\documentclass[12pt]{article}
\usepackage[tmargin=1in,bmargin=1in,lmargin=1.25in,rmargin=1.25in]{geometry}

\usepackage{setspace}
\usepackage{csquotes}
\usepackage{authblk}
\usepackage{kpfonts}
\usepackage{amsmath}
\usepackage{amsfonts}
\usepackage{stmaryrd}
\usepackage{dsfont}
\usepackage{graphicx}
\usepackage{comment}
\usepackage[title]{appendix}
\usepackage[font=small,labelfont=bf]{caption}
\usepackage{lipsum}
\usepackage{mwe}
\graphicspath{ {images/} }

\title{ A Partial Defense of Algebraic Relationalism}
\date{}
\author{Lu Chen\thanks{University of Southern California. Email: chen.l@usc.edu}}

\begin{document}
	\maketitle
	
	\begin{abstract}
I defend  algebraicism, according to which physical fields can be understood in terms of their structural relations without reference to a spacetime manifold, as a genuine relationalist view against the conventional wisdom that it is equivalent to substantivalism, according to which spacetime exists fundamentally. I criticize the standard version of algebraicism that is considered equivalent to substantivalism. Furthermore, I present alternative examples of algebraicism that better implement relationalism with their own conceptual advantages over substantivalism or its standard algebraic counterpart.
	\end{abstract}

\tableofcontents

\section{Introduction}

Does space (or spacetime) exist independently of material bodies? According to \textit{substantivalism}, the answer is yes, while \textit{relationalism} says no---only material bodies exist fundamentally. The debate between these two views has become largely stifled since the advent of general relativity. In the standard formulation of general relativity, spacetime is conceptualized as a differentiable manifold equipped with a metric field, which determines the distances between any two points that are connected by a path.\footnote{While there are many variants of general relativity that do not postulate a primitive metric field, these variants do not make a difference to our discussion---they all appeal to some entity that determines the geometric feature of spacetime. See Krasnov [2020].} Spacetime models that involve a manifold and metric field have since become prominent in physics. Once we endorse these entities realistically, there is little room for relationalism (see Maudlin [1993]). While many reconstrue the metric field as a matter field on a par with other physical fields, it is less standard to reject \textit{manifold substantivalism}, the view that says an amorphous spacetime manifold exists fundamentally and independently of material bodies (or fields). 

One way to revive the debate is to look at alternatives to the manifold-theoretic approach to physics.\footnote{There are other ways of reviving the debate. See, for example, Belot [1999], Pooley [2013], and North [2018].} Let's call the manifold-based formalism \textit{manifoldism}.\footnote{My use of `manifoldism' will be exchangeable with that of `manifold substantivalism', except that when they are initially introduced and when the difference matters, the former emphasizes a physical framework, while the latter is a metaphysical view based on this framework.} As an alternative, Geroch ([1972]) proposed an algebraic reformulation of general relativity, which dispenses with any reference to a manifold. In this reformulation, physical fields are not characterized by maps from the spacetime manifold, but by their structural relations with each other. These relations are characterized by an algebraic structure called `Einstein algebra', on which various notions relevant to general relativity are redefined. Let's call such an algebra-based, manifold-free formalism \textit{algebraicism}. Understood as such, algebraicism naturally leads to a relationalist interpretation: physical fields exist fundamentally without an underlying spacetime. Indeed, Earman ([1977],[1989]) introduced Geroch's mathematical formalism as an implementation of relationalism---call it \textit{algebraic relationalism}.\footnote{Again, `algebracism' and `algebraic relationalism' are interchangeable in this paper, except that when they are introduced and when the difference matters, the former emphasizes a formal framework of physics and the latter a metaphysical interpretation.} He wrote ([1989], p.192):\footnote{\label{Leibniz} Earman used the term `Leibniz algebra'  instead of `Einstein algebra'. For simplicity, I will stick with the latter terminology throughout.} 

\begin{quote}
	Thus, it is open to take the [Einstein algebra] as giving a direct characterization of physical reality… This provides one plausible reading, I think, of Einstein's idea that space-time, M, ‘does not claim an existence of its own, but only as a structural quality of the field’.
\end{quote}
It is worth mentioning that Geroch did not introduce algebraicism with relationalism in mind. Rather, he worried that the manifold-theoretic approach to physics may be an obstacle for making progress in quantum physics (`a quantum theory of gravitation...will suggest a smearing out of events', p.271). In other words, algebraicism is also motivated independently of typical relationalist considerations.

However, ever since its introduction, algebraic relationalism has been viewed or criticized as a disguised substantivalist view.\footnote{A different line of arguments can be found in Bain [2006], where he argued that the formalism of Einstein algebras supports spacetime structuralism which he saw as an alternative to both substantivalism and relationalism.} Even Earman ([1989], p.193) himself complained about the `ghost of substantival space-time' in Einstein algebras. Later, Rynasiewicz ([1992], p.573) influentially argued: `For the category of topological spaces of interest in spacetime physics, the algebraic program is completely equivalent to the spacetime approach. The path of spacetime algebraicism thus leads right back to substantivalism.' Rosenstock et al ([2015], p.310), while endorsing algebraicism as a relationalist view, concluded that `a suitably “relationist” theory, once one spells it out in sufficient detail, is equivalent to the “substantivalist” theory.' This sentiment is again echoed in more recent work (for example, see Linnemann and Salimkhani [2021]).

Why is algebraicism considered substantivalism in disguise? Rosenstock et al ([2015]), in particular, demonstrated the equivalence between algebraicism and manifoldism within the context of general relativity, where theoretical equivalence is understood in terms of `categorical duality' (see Halvolson [2012], Weatherall [2016]). As a result, insofar as manifoldism commits to a substantival spacetime, algebraicism does so too as merely a different way of conceptualizing the same reality. 

Against this conventional wisdom, I will argue that algebraic relationalism is a genuine relationalist view that is alternative to substantivalism, and has some distinct advantages over the latter. Note that the aim of the paper is precisely that---\emph{nothing more}. In particular, I will not give a complete cost-benefit analysis of algebraic relationalism in comparison with other brands of substantivalism and relationalism.\footnote{Of course, in order to defend algebraic relationalism as a relationalist view, I will still address some immediate concerns about algebraic relationalism independently of the equivalence claim, such as regarding its ontology. But a comprehensive discussion of all related issues is beyond the scope of this paper.} I think this moderate goal is justified because of the prevalence of this conventional wisdom and its ramifications.\footnote{Even when one does not view algebraic relationalism as a disguised substantivalist view, one may still attach other significance to the equivalence claim. For example, Bain [2006] used the equivalence claim to argue that we should endorse spacetime structure rather than its particular realization through manifold points or scalar fields.}

More detailedly, I will argue that the formalism of Einstein algebras that the philosophical literature appeals to is a very specific, non-representative implementation of algebraicism, which is not justified by empirical and theoretical considerations from a relationalist point of view.  So, even if we grant that this particular algebraic implementation of relationalism is equivalent to manifold substantivalism, this is not a particularly interesting conclusion and begs the question against relationalism.



To illustrate why algebraicism is not equivalent to manifoldism, I will appeal to three different examples of algebraicism in the literature. The first, which is based on synthetic differential geometry, shows that in the algebraic framework, we have a natural way to define an algebra that can be heuristically understood as an `infinitesimal region', which is \textit{not} intrinsically definable in manifoldism or more generally, a geometric-theoretic framework (see also Lawvere [1980], Kock [1981], Chen [2022]). Such an `infinitesimal region' can facilitate a natural interpretation of vectorial quantities.  This example is particularly rigorous and illuminating for showing why algebraicism is not equivalent to manifoldism, even within the context of general relativity. Indeed it provides a clear counterexample to the criterion of equivalence based on categorical duality.

The second involves `multi-field algebras' proposed by  Chen and Fritz ([2021]) that promise to  posit only physical fields acknowledged by standard physics, without any `ghost' field that plays the role of spacetime.\footnote{These authors call the algebraic structure in question simply `field algebra'. As will become clear later, I take `multi-field algebra' as more descriptive such a structure. } In contrast, the standard algebraic formalism makes essential reference to a scalar field, which may not be among our fundamental physical fields. This example shows that a more ontologically perspicuous algebraic formulation of relationalism is not equivalent to manifold substantivalism.

The third example, which comes from non-commutative geometry, shows how we can generalize manifoldism by lifting the requirements of commutativity for algebras. Such algebras do not correspond to any traditional geometric structures, but can still constitute a framework for physics, and quantum theory in particular (for example, see Connes and Marcolli [2008]; also see Huggett et al [2021]). Recall that one motivation for algebraicism suggested by Geroch is its potential role in advancing quantum physics. Once again, this example shows that a physically motivated algebraicism is not equivalent to manifoldism.

Finally, a word about the technical presupposition of the paper: I intend the philosophical discussion to be accessible to general readers who are interested in the substantivalism-relationalism debate and in particular how the algebraic approach fits into this debate. Although the paper is in part highly technical (which, for example, presupposes familarity with category theory), I shall emphasize that an understanding of all or most technical details is not a presupposition for grasping the philosophical moral. For readers with the relevant background, I intend the technical details that directly serve the philosophical claims to be rigorous and self-contained.


		\section{The Equivalence Claim}
		
		What is algebraicism? 		Let us start with a simple world described in geometric terms: spacetime together with a scalar field. Possible configurations of the scalar field are represented by smooth real-valued functions on the manifold (`smooth' means indefinitely differentiable). Notice that these configurations bear structural relations with each other. For example, we can add or multiply two functions to get another one. The basic idea of algebraicism is that we can understand them just in terms of these relations alone, without reference to an underlying manifold. Furthermore, these relations can be formulated algebraically: the field configurations form an algebraic structure defined by these relations. In this approach, instead of taking these field configurations as functions from the manifold to real numbers, we can re-conceptualize them as simple elements of the algebra. 
		
			The crucial contribution of Geroch ([1972]) is to show that based on particular implementation of algebraicism called `Einstein algebras' we can redefine all notions of differential geometry that allows us to do physics up to general relativity. 	An Einstein algebra is defined as consisting of a commutative ring with a subring isomorphic to $\mathbb{R}$ and an algebraic metric defined on the ring that satisfies several conditions. We can write down the Einstein field equation and its solution in terms of Einstein algebras without reference to a manifold.  As a simple example, we can reconstruct spacetime points as follows. Let  $C^\infty(M)$ denote an Einstein algebra isomorphic to all smooth real-valued functions on manifold $M$, and $\hat{g}$ an algebraic metric on it.\footnote{For simplicity, I will henceforth ignore the algebraic formulation of the metric field, which is not crucial for my arguments.} In this approach, a model of this world is not formulated as $\langle M,g, \phi\rangle$, where the metric $g$ and the scalar $\phi$ are solutions to relevant dynamical equations, but rather as $\langle C^\infty(M), \hat{g}, \hat{\phi}\rangle$, where $\hat{\phi}$ is simply an element of $C^\infty(M)$ corresponding to $\phi$.
		
		To interpret the algebraic models realistically, we include all elements of $C^\infty(M)$ into our ontology and all algebraic structures into our ideology. Instead of claiming that manifold exists fundamentally, we claim that all scalar field configurations exist fundamentally without a manifold. This can invoke `the incredulous modal stare': we now have to posit the existence of infinitely many merely possible field configurations as opposed to just one configuration that is instantiated by the actual world (see, for example, Butterfield [1989]). While I will not delve into a thorough defense, I shall point out that this worry is not as severe as it may seem. Even in standard physics, merely possible field configurations play a significant role. For example, a dynamical law governing physical fields typically quantifies over all possible field configurations, not just those that are concretely instantiated.\footnote{It is worth mentioning that some people, such as Chen E.[2021], argued that initial conditions also constitute a law of nature, which pins down a unique world history.} If one is \textit{primitivist} about laws of nature, namely that laws are brute fact irreducible to spatiotemporal patterns of the world, one needs to assign some ontological status to the possible field configurations that the laws refer to. Moreover, if we turn to quantum  theory, the path integral formulation explicitly involves summing over all possible field configurations, which all have salient physical significance (e.g., have amplitudes that contribute to the observables). It is also worth noting that it is common for relationalism to appeal to merely possible material bodies or processes (for example, see Belot [1999]). While this is not an attractive feature of the view, it is not a sufficient ground for rejecting the view or trumping other considerations. Finally, when positing merely possible field configurations, we do not need to construe them as mysterious or sui generis abstracta. We can understand each configuration as a state of the field, and approach the ontology of states through high-order metaphysics that allow for higher-order entities like states, a framework motivated independently of algebraicism (see Bacon [2022]). 	There are remaining ontological issues pertaining to the particular formalism of scalar fields in Einstein algebras, which I will discuss and remedy in Section 4.2.\footnote{\label{Pl} There are similar worries about the plenitude of algebraic relations in the ideology of algebraicism that do not necessarily have physical significance.  I shall first note that it is possible to only appeal to the algebraic operators that are used in physics, such as addition and multiplication of field configurations. But to address this worry satisfactorily, we will need an in-depth discussion of `physical significance' (see Curiel [2016]).}

		Now, let's turn to why manifoldism and algebraicism are perceived as equivalent in the literature. I will first explain Rynasiewicz's ([1992]) argument that we can reconstruct manifold-theoretic structures under algebraicism.\footnote{Rynasiewicz's overall claim is that algebraicism does not help solving the hole argument that troubles substantivalism. I think Rynasiewicz is correct on that, even when it applies to the more general algebraic approach. But I will not focus on the hole argument in this paper. While the hole argument is often used as a motivation for relationalism, it is not the only motivation, nor is it a particularly strong one, given that there are many proposed solutions to it that do not dispense with fundamental spacetime structures.} Then I will present Rosenstock et al's ([2015]) more formal argument for their equivalence claim with a precise definition of `equivalence' as categorical duality.

	To argue that  an Einstein algebra  is `just a substantival model in disguise' ([1992], p.583), Rynasiewicz showed that the spacetime structures in manifoldism can be reconstructed in algebraicism.\footnote{Note that Rynasiewicz called it `Leibniz algebra' following Earman [1977] and [1989]. See Footnote \ref{Leibniz}.}  An `ideal' of an algebra is a subset of it that can be consistently identified with zero, and therefore we can use an ideal to define a quotient algebra by `collapsing' elements of the original algebra that at most disagree with respect to the ideal. A `maximal' ideal is a largest ideal short of being the entire algebra. For example, all smooth functions that vanish at point zero is a maximal ideal of $C^\infty(M)$---call it $I_0$. It defines a smallest non-zero quotient algebra (denoted by `$C(M)/I_0$') that is isomorphic to $\mathbb{R}$ by collapsing all functions that agree on point zero. Now, as  Rynasiewicz pointed out (which is also a basic fact in algebraic geometry), such maximal ideals can be conceived as spacetime points because there is a one-to-one correspondence between points of $M$ and maximal ideals of $C^\infty(M)$ that preserves the topology (the collection of all maximal ideals each of which contains an element of $C^\infty(M)$ form a basis of closed sets for the topology defined of the set of all maximal ideals of $C^\infty(M$); see Rynasiewicz [1992], pp.583-4 for more details).\footnote{A technical caveat: without qualifications, the notion of `maximal ideal' is more general than that of a point. In the case of $C^\infty(M)$, it can be shown that there is a maximal ideal that contains all functions whose supports (say) have upper bounds. This would correspond to `a point in the positive infinity' but there is no such point in the standard real line. The one-to-one correspondence applies when we rule out such cases.} 	This shall work as an illustration for how we can construct spacetime structures in manifoldism on the basis of an Einstein algebra, and I will stop short of presenting more technical details.
		
		We have the result that the maximal ideals of $C^\infty(M)$ behave just like the points in $M$. Does this mean that spacetime points have been reconstructed in the algebraic framework? How should we think of those maximal ideals? Bain ([2003]), among others, suggested that we can adopt what he called `function literalism', according to which a maximal ideal is literally a plurality of smooth functions (taken as fundamental in our ontology), not a spacetime point. While this interpretation is coherent per se, claiming the resulting ontology free of points conflicts with a structuralist spirit that I wish to preserve. Consider again the manifold $M$. In the usual mathematical practice, it does not matter how we represent the elements of $M$ but only how we define the topological and differential structure on the elements. Thus, the elements of $M$ can just as well be the maximal ideals of $C^\infty(M)$! Since these maximal ideals are indeed included in the ontology of algebraicism, by this construction $M$ is also included in the ontology. Thus I agree with Rynasiewicz that Einstein algebras give rise to spacetime points. 
		
		Rosenstock et al ([2015]) gave a rigorous demonstration that Einstein algebras as models for spacetime theories are equivalent to manifold-theoretic models, with `equivalence' being precisely defined as categorical duality (see Halvolson [2012], Weatherall [2016]). Formulating manifoldism and algebraicism in category theory is very useful for extracting and comparing their structural features, since it concerns the relations between manifolds or algebras rather than their internal structures. Rosenstock et al shows that the category of manifolds \textbf{Man}, with manifolds as objects and smooth maps between them as morphisms, is dual to the category of `smooth algebras' \textbf{Alg}, with smooth algebras as objects and homomorphisms between them as morphisms. `Smooth algebras' are Einstein algebras that satisfy a few additional conditions that I will discuss in Section 3, but for now the details are unimportant. This duality means that there is a one-to-one correspondence between the objects in \textbf{Man} and \textbf{Alg}, namely between smooth manifolds and smooth algebras, that preserves all their relations except the directions of the morphisms are reversed.\footnote{More rigorously, two categories are dual or called `dually equivalent' if there are two contravariant functors that are inverse to each other up to isomorphism (namely, their compositions are naturally isomorphic to identity functors). The definition for the equivalence of categories is the same except that the `contravariant functors' in question should instead be `covariant'. This is a weaker condition than isomorphism: for example, a category with one object and a trivial identity arrow can be (dually) equivalent to a category with two objects with two arrows between them. In other words, equivalent categories can have different numbers of isomorphic copies.} For example, a smooth map $\phi$ from $M$ to $N$ corresponds to a homomorphism from $C^\infty(N)$ to $C^\infty(M)$, which maps a smooth function $f$ on $N$ to a smooth function on $M$ by simply composing $f$ with $\phi$ itself. This duality can also be extended to two categories containing additional structures to  \textbf{Man} and \textbf{Alg}.\footnote{Let \textbf{GR} be the category of relativistic spacetimes usually modelled by $\langle M, g\rangle$ with the arrows being the isometries between them (an isometry is a smooth map that preserves the metric structure $g$). These objects are solutions to Einstein's field equations, or in other words, models for general relativity. Let \textbf{EA} be the category of `Einstein algebras' with the arrows being homomorphisms between them. Rosenstock et al defined `Einstein algebras' to be pairs of smooth algebras and algebraically defined metrics $\langle A,\hat{g}\rangle$ (the detail is different from Geroch's, but that does not matter for our discussion). The authors also proved that \textbf{GR} and \textbf{EA} are dual categories. This implies that for every manifold-theoretic model of general relativity, we automatically have an Einstein algebra as an algebraic model.} Based on this, Rosenstock et al concluded that manifold-theoretic and algebraic formulations of general relativity are equivalent, which I will argue against in the next section.
		
		
		
		
		It might be worth emphasizing that Rynasiewicz's ([1992]) conclusion that algebraicism does not help solve the infamous hole argument is independent from the equivalence between manifoldism and algebraicism that I will argue against. Whether or not they are equivalent, algebraicism has the analogous problem of manifoldism that algebraic models can be isomorphic but distinct (in the same sense that spacetime models can be isomorphic but distinct), giving rise to its own version of the hole argument.  Like many authors in the literature, I believe that the key to solve or dissolve the  argument lies elsewhere beyond the scope of this paper.\footnote{I argue elsewhere in favor of the approach that appeals to a new logico-mathematical foundation, the Univalent Foundations.}

	\section{Against Einstein Algebras}
		
		It has been demonstrated that the formalism of Einstein algebras is, in a sense, equivalent to manifoldism. However, this does not mean that algebraicism in general is equivalent to manifoldism, because this particular implementation of algebraicism is not representative of algebraicism.  I will argue that this formalism is neither a motivated nor an interesting implementation of relationalism in that it involves unjustified and artificial constraints, and ontologically privileges the scalar field over other matter fields. 
		
		\subsection{Warding off confusions}
		
		But first, I think it important to respond to some common misunderstandings of algebraicism that lead some people to think that Einstein algebras are more or less representative of algebraicism in general, and that algebraicism is more or less equivalent to manifoldism, or otherwise fails to truly dispense with manifolds.\footnote{I encounter the following confusions in many correspondences. My impression is that such confusions are very common among those who do not work heavily on this topic and hope this discussion will be of value to everyone who is interested in algebraicism.} 
		
\begin{quote}
		\emph{Confusion 1: Manifolds are used in the construction of algebraic structures and therefore presumed.}
\end{quote}
	To an expert in the literature, this confusion may sound too naive, but it is still important  to address. This is a very natural confusion. When we talk about algebraic structures, we seem to constantly refer to manifolds. For example, Einstein algebras are denoted by $C^\infty(M)$, with $M$ being a smooth manifold. Even if we claim that the elements of $C^\infty(M)$ are considered primitive, it may seem that we are still using a manifold in their representations. 
	To respond, we just need to acknowledge that algebraicism can thoroughly dispense with manifolds in describing algebraic structures. In Geroch's formalism, although we start with smooth functions on the manifold, we reconceptualize  the structure as simply a commutative ring with a subring isomorphic to the ring of real numbers, which makes no reference to manifolds. Other notions required for physics are defined solely on the algebraic representation of $C^\infty(M)$ without  references to manifolds. To see that we have really `forgotten' about the manifold in the algebraic formalism, it would be helpful to note that it is a highly non-trivial task to conceptualize a manifold-theoretic structure algebraically, as I will explain soon.
		
		\begin{quote}
		\label{scalar}	\emph{Confusion 2. Smooth functions (or in physical term, scalar field configurations) involved in Einstein algebras are accidental to the formalism. We can readily formulate algebraicism using other fields of interest. }
		\end{quote}
		This is a common reaction to the complaint that the smooth functions involved in Einstein algebras are not physically real (since we do not have a physical field that is a real-valued scalar field; see Chen and Fritz [2021]). The thought is that this complaint is moot because appealing to smooth functions or a scalar field is just a toy example rather than committal, since the case of a scalar field is the simplest. To extend algebraicism to physically more realistic situations,  we can replace the scalar field by other fields of choice. While this reaction is very understandable, it fails to recognize that smooth functions are built into the basic framework of Einstein algebras on which every other notion is defined, and cannot be readily replaced by any other fields. To illustrate this, let's consider a vector field instead.
		
		 Consider the set of all possible configurations of a vector field on the manifold $M$. Given that the algebraic formulation of the scalar field encodes all the information about $M$, one might expect that we can also find an algebraic formulation of the vector field that contains all the information about $M$, especially since the vector field contains more information than the scalar field (to each spacetime point, the scalar field assigns a real value while a vector field assigns both a real value and a direction). However, this is not true. Unlike smooth functions, vectorial functions do not form a commutative algebra and therefore fails to satisfy the basic definition of Einstein algebras.\footnote{Tensor fields, which play a particular important role in physics, face the same problem as vector fields. We can add two tensor field configurations together to get another, and we can multiply one by a scalar field configuration, but there is no suitable multiplication operation that map two tensors to another in order to define a sufficiently rich algebraic structure. Although the tensor product operation takes two tensors to another tensor, they are of different ranks, and therefore the algebraic structure defined must not consist of only the configurations of a particular tensor field.   } The operation in question needs to map two vectorial functions to a vectorial function, but the usual multiplication for vectors is between a vector and a real number. The inner product between two vectors does not work either because it maps two vectors to a real number rather than a vector. The cross product is also infeasible because it is not associative nor commutative. If we let vectorial functions form just an abelian group equipped with addition, then the algebraic structure is more impoverished than that of the scalar field. Without the latter structure,  other notions in the Einstein-algebra formalism cannot be defined.  In relation with the first confusion, note that the manifold occurring in $V(M)$ is inconsequential after we reconceptualize the structure algebraically---the manifold is not automatically smuggled into the algebraic structure.

This example, I hope, helps illustrate why the scalar field plays a crucial role in the formalism of Einstein algebras, which is not readily replaceable by any other field. This does not mean that algebraicism must resort to scalar fields, but that dispensing with the scalar field would amount to a substantial revision of Geroch's formalism based on Einstein algebras. 

\subsection{First objection}
I will now turn to my claim that Einstein algebras featured in the equivalence claim are artificial and unmotivated. To ensure that we can fully recover all standard notions for general relativity, Rosenstock et al ([2015]) required an Einstein algebra to be `geometric', `complete', and `smooth' so as to be equivalent to a manifold. I will argue that each of them lacks an intrinsic justification. 

It is worth first highlighting that the conditions are not posited in Geroch ([1972]) where he argued that we can theorize general relativity using Einstein algebras. Thus, they are not motivated by the need of physics. Nor does it seem to be based on any consideration other than proving the equivalence claim itself.

 \label{nilpotent} First, consider geometricity. An algebra $A$ is called `geometric' if no non-zero elements lie in the kernel of all homomorphisms from $A$ to $\mathbb{R}$ (the kernel of a homomorphism is the set of elements that are mapped to zero by the homomorphism). In other words, for any non-zero element of such an algebra, we can find a homomorphism that takes it to a non-zero real numbers. This would rule out any algebra that contains `nilpotent' elements which square to zero (call such algebras `nilpotent algebras'), since homomorphism must preserve the multiplicative relations.  As an example, this would rule out the set of affine functions (with the form of $f(x)=a+bx$),  equipped with the `affine approximation' of the usual multiplication.\footnote{For any affine functions $f,g$, if $f\cdot g$ denotes multiplication in the usual sense, then the multiplication of $f$ and $g$ in the algebra of affine functions is $(f\cdot_ag)(x)=(f\cdot g)(0)+(f\cdot g)'(0)x$. Now, consider $f(x)=bx$. If we square this element, we obtain $f^2(x)=b^2x^2=0$.} Call this algebra  $\mathfrak{A}$.

Why should we rule these out for algebraicism? No rationale is given by the authors other than the apparent reason that including such algebras would falsify the equivalence claim, which is what the authors set out to demonstrate. In terms of categorical duality, nilpotent algebras are not dual to any manifold. This can be shown by the following reasoning. There is in fact only one homomorphism from $\mathfrak{A}$ to $\mathbb{R}$, the one that maps $f(x)=a+bx$ to $a\in\mathbb{R}$ (that $f(x)=bx$ is mapped to zero is a special case).\footnote{$\phi$ be a homomorphism from $\mathfrak{A}$ to $\mathbb{R}$. For any $a\in\mathbb{R}$, $\phi$ must map constant function $g=a$ to $a$ since it preserves zero element and addition. For any $h(x)=bx$, $\phi$ must map it to 0. Then, $\phi$ must map $f(x)=a+bx$ to $a$ to preserve addition.} In the dual category, the corresponding claim is that there is only one smooth map from a point (zero-dimensional manifold) to the dual object of $\mathfrak{A}$. This means that this dual object has only one point (an object in the geometric category contains as many points as the embedding maps from a point), but it is not identical to a point! This immediately tells us that this cannot be any standard manifold or indeed any point-set structures. While such algebras correspond to no standard geometric objects, there is no obvious reason to rule them out from an algebraic point of view. Indeed, I will give reasons for keeping them in the next section. To summarize the upshot:

\begin{quote}
	\emph{Nilpotent algebras correspond to no point-set regions.}
\end{quote}

Secondly, consider the requirement of `completeness'. This is a requirement on what is called the `restriction homomorphism'. Here is how it is defined in Rosenstock et al. Let $S$ be a set of homomorphisms from an algebra $A$ to $\mathbb{R}$. $A_{|S}$ consists of elements of $A$---in this context, smooth functions---restricted to $S$, which is also an algebra. More precisely, Rosenstock et al ([2015], p.311) defined $A_{|S}$ as consisting of smooth functions $f$ on $S$ such that for every point $x\in S$ there is a neighbourhood $O$ of $x$ in $S$ on which there is a function in $A$ that agrees with $f$.  The homomorphism from $A$ to $A_{|S}$ that takes $f\in A$ to $f_{|S}\in A_{|S}$ is called the `restriction homomorphism'. $A$ is said to be `complete' if its restriction homomorphism is surjective with regard to the set of \textit{all} homomorphisms from $A$ to $\mathbb{R}$ (pp.311-2). Here, to make sense of the restriction homomorphism, we already need to think geometrically, namely to consider $S$ as a set of points in the dual space of $A$ so that we can `restrict' $f$ to $S$.\footnote{The explicit definite of $A_{|S}$ given is not even fully unpacked in algebraic terms, but rather stated in geometric terms in the dual category. For example, it involves functions defined on $S$, points in $S$, and local neighbourhoods of points in $S$. In all these locutions, $S$ is conceptualized as a space in the dual category rather than its algebraic definition, namely a set of homomorphisms from $A$ to $\mathbb{R}$. In other words, the constraint is not even spelled out algebraically, let alone motivated algebraically.} We may conclude  not only that the condition is unmotivated, but also the following:

\begin{quote}
	\emph{To require algebras to be complete, we make essential reference to geometric objects, which suggests this requirement as a disguised geometric discourse.}
\end{quote}

Finally, the condition of smoothness suffers from the same problem. A geometric and complete algebra is called `smooth' if there is a countable open covering on the set of points in the dual object of $A$ that satisfies some further constraints. Here, the geometric concepts are directly invoked---they are not translated to algebraic terms, let alone justified algebraically. If we take this method to its extreme, we would simply be stipulating that Einstein algebras (or rather, smooth algebras) should be such that they constitute a category that is dual to the category of manifolds, which is patently unsatisfying.\footnote{This is not necessarily a serious criticism of Rosenstock et al's work, since their goal is to ensure all the standard notions of general relativity can be recovered in algebraicism.} 

To take stock, the equivalence claim requires us to impose various constraints on Einstein algebras that are artificial and not independently motivated. 

\subsection{Second objection}

My next objection is that the scalar field is built into the formalism of Einstein algebras (see \ref{scalar}). This is problematic for algebraicism as a relationalist view because the scalar field is essentially a surrogate manifold that represents spacetime. 

 First of all, there is no actual real-valued scalar field at the fundamental level acknowledged by current physics.\footnote{The closest to this is the Higgs field, which is a complex-number-squared-valued scalar field. Admittedly, Einstein algebra can be realized by such a  scalar field, it would be strange to interpret the fundamental algebraic structure as consisting of the Higgs field.} But even if there is, it is possible there isn't. But the formalism of Einstein algebras requires that a scalar field necessarily exists. Furthermore, this formalism privileges the field ontologically over all other matter fields. Since all physically relevant structures are to be defined in terms of the algebra $C^\infty(M)$, non-scalar fields such as the electromagnetic field would have to be defined in terms of the scalar field. But it is implausible that the electromagnetic field depends on and presupposes the existence of the scalar field. All these indicate that the scalar field behaves just like spacetime in substantivalism. If algebraic relationalists are unhappy with the ghostly arena of physical on-goings, they should be equally unhappy about a ghostly background scalar field that must be posited in addition to those acknowledged by standard physics. (This also motivates many people to consider algebraicism a substantivalist view.) 
 
 Note that there are many nonfundamental scalar fields in physics, such as the mass-density of a matter field. But these do not help much with the interpretative issue, since it is implausible to ontologically privilege such a scalar field over other fields that are standardly considered to be more fundamental.

What if we refrain from interpreting $C^\infty(M)$ realistically, as Menon ([2019]) suggested? That is, $C^\infty(M)$ is considered a mathematical abstracta rather than part of physical reality. But this strategy has the same problem as (for example) interpreting manifold abstractly in the relativistic physics based on standard differential geometry. As Norton ([2008]) pointed out, without a manifold, the coordinate system on which the physical fields are defined should be interpreted realistically, for otherwise we could not account for the spacetime coincidence of fields. Analogously, elements of $C^\infty(M)$ are invoked to describe the spacetime coincidence of fields.  For example, consider two vector fields which have non-zero values at some points and zero elsewhere. How do we describe whether the non-zero values are located at the same region or not? In the coordinate-system approach, we simply compare their values at each point in a coordinate system. Analogously, in the algebraic framework, we compare how they act on  $C^\infty(M)$. In the formalism of Einstein algebras, a vector field is a collection of maps from $C^\infty(M)$ to $C^\infty(M)$ called `derivations' (Geroch [1972], p.272). The non-zero values of the two fields $\Psi,\Phi$ locate at the same region just in case for all $f\in C^\infty(M)$, $f\Psi=0\leftrightarrow f\Phi=0$, where $f\Psi$ is defined by $(f\Psi)(g)=f\Psi(g)$ for all $g\in C^\infty(M)$. The idea is that, if the regions that the two fields have non-zero values are the same, then we can `detect' this by multiplying both fields by all functions that are zero at that region and see if they behave the same. Thus, if algebraicists are discontent with reading manifold abstractly in the standard framework, then the analogous role played by $C^\infty(M)$ should be equally troubling.

\subsection{Third Objection}
As I mentioned in the beginning, one reason to get rid of manifolds is due to the need for new physics. An algebraicism equivalent to manifoldism would not contribute to this agenda. Thus, if we aim for an algebraicism that opens up new possibilities in physics, Einstein algebras are not sufficiently interesting.

\paragraph{}
To take stock, the formalism of Einstein algebras is not representative of algebraicism as a relationalist view. It just demonstrates is that algebraicism has at least the expressive power of manifoldism in that we can develop an algebraic foundation equivalent to the latter.
Against the conventional wisdom, the philosophical lessons learned from Einstein algebras can apply only very limitedly to algebraicism or algebraic relationalism in general.

\section{Cases for Algebraicism}

I have argued against the formalism of Einstein algebras. To make a positive case for algebraicism, I will give three less-discussed examples of algebraicism from the literature that are  not equivalent to manifoldism or substantivalism and have their distinct advantages over the latter.  Note that, however, they are all preliminary frameworks and do not address the objections in the last section \textit{all at once}.  But to argue against algebraicism solely on this ground is cheap (which is perhaps analogous to arguing against research programs in quantum gravity, as none of them has achieved the results they aim for). Instead, I hope these examples will, in addition to preparing a better ground for philosophical discussion, inspire readers to develop algebraicism further.

\subsection{Nilpotent algebras}

As mentioned in Section 3.2, if we include `non-geometric' algebras, the resulting category is not equivalent to the category of smooth manifolds. The resulting category is instead a part of a framework called `synthetic differential geometry' (Lawvere [1980], Kock [1981]). To understand this framework, it would be helpful to understand the significance of the `non-geometric' or nilpotent algebras.

First, it is worth emphasizing that nilpotent algebras are \textit{naturally} included in the category of algebras for algebraicism. Let me elaborate. While a field configuration spread over all of spacetime can be conceptualized as a basic entity in algebraicism, it is clear that we still need to talk about a part of a field.\footnote{I will put aside the issue of what is more fundamental: the whole field or its parts. Both options are feasible without further arguments.} This is facilitated by the notion of ideals or quotients (see Section 2). To talk about parts of the fields in $C^\infty(M)$, we appeal to its quotient algebras.\footnote{For example, Let $I$ be the set of smooth function that vanish at $[0,1]$. Then $C(\mathbb{R})/I$ is isomorphic to the set of smooth functions on $[0,1]$ (since we identify all functions that agree on $[0,1]$).} The nilpotent algebra $\mathcal{A}$ consisting of affine functions is precisely one such quotient algebra and behaves just as well as other quotients.\footnote{$\mathfrak{A}$ is isomorphic to the quotient algebra $C^\infty(\mathbb{R})/I_\Delta$, with the ideal $I_\Delta$ consisting of all smooth functions that have zero value and zero derivative at point zero. } Their parthood relation is represented by the quotient map from $C^\infty(M)$ to $\mathfrak{A}$, which corresponds to an embedding map between the dual object of $\mathfrak{A}$ to $M$.


But what is the dual object of $\mathfrak{A}$---call it `$\Delta$' (following Bell [1998])? 
We already know that this is not a manifold (Section \ref{nilpotent}). We can consider it as an infinitesimal neighborhood around a point since it is properly contained in every finite region around the point and contains information about the derivative of a function.\footnote{A finite interval region $[a,b]$ corresponds to the ideal of all smooth functions that vanish on the interval $[a,b]$. It is easy to confirm that all infinitesimal regions around a point are contained in all the finite intervals around that point: if a function vanishes at an interval, then its $i$-th derivative at a point properly within the interval has to be zero for any $i$.  Another way of seeing $\Delta$ as infinitesimal is that the corresponding quotient algebra throws out all information of a smooth function other than its value and derivatives at a point. }
The derivative of a function at a point is just the slope of the `infinitesimal' part of the function around that point (that is, the restriction of the function to $\Delta$).\footnote{\label{derivative} More concretely, suppose $f$ is a smooth function from $\mathbb{R}$ to $\mathbb{R}$. To know what its derivative is at a point, we just need to restrict $f$ to the infinitesimal region around the point, which simply means composing $f$ with the relevant map from $\Delta$ to $\mathbb{R}$. We can show how this works from the algebraic point of view. $f$ corresponds to a homomorphism $\phi$ from $C^\infty(\mathbb{R})$ to itself. A map from $\Delta$ to $\mathbb{R}$ corresponds to a homomorphism $\psi$ from $C^\infty(\mathbb{R})$ to $\mathfrak{A}$, which takes any $h\in C^\infty(\mathbb{R})$ to (for example) $g(x)=h(0)+h'(0)x$. So the composition of them will take any $h\in C^\infty(\mathbb{R})$ to $g(x)=\phi (h)(0)+\phi(h)'(0)x$. Then we can use the projection map to extract the first-order coefficient which is the desired derivative of $f$ at zero.} As it turns out, there is no way to analytically describe $\Delta$ as a point set within classical logic.  In Section \ref{nilpotent}, I have mentioned that $\Delta$ contains only one point (since there is only one homomorphism from its dual object $\mathfrak{A}$ to $\mathbb{R}$, the dual algebra of a point). Yet $\Delta$ is not a point or the singleton of a point, which cannot be made intelligible within classical set theory (see also Moerdijk and Reyes [1991], Bell [2008], Hellman [2006] for detailed discussions).\footnote{As Moerdijk and Reyes ([1991]) have shown, the category of manifold augmented by nilpotent objects constitute a model for smooth infinitesimal analysis, where $\Delta$ has the following two classically contradictory features: (1) there are no points in $\Delta$ that are not identical to point zero; (2) it is not the case that all points in $\Delta$ are identical to zero (see Bell [2008] for an exposition of smooth infinitesimal analysis). Hellman ([2006]) argued that we cannot interpret these features classically.} 

But why should we consider such a theory? To begin with, the simplicity and naturalness of obtaining derivatives are considered an advantage of the algebraic framework among algebraic geometricists. But a more important conceptual advantage is that this provides a simple way of interpreting vectorial quantities metaphysically (Chen [2022]; see also Weatherson [2006]). In the standard framework, it is puzzling whether a vector (such as an electric field value) at a spatial point is intrinsic to a point or not. It would be bizarre for an extensionless and directionless point to contain a quantity with spatial direction, so a vector should be considered extrinsic to a point. But then we would be endorse a metaphysics that commits to fundamentally extrinsic properties, which is standardly considered undesirable. The problem multiplies when we consider the nature of tangent spaces in curved spacetime. With infinitesimal regions, we can simply say that a vector is intrinsic to an infinitesimal region, and can be understood as an infinitesimal part of a scalar field or a vector field. 

Finally, I want to point out a more general lesson from this case, although a more detailed discussion needs a separate treatise: the formal duality between the geometric category and the algebraic category does not mean that the corresponding geometric theory and the algebraic theory are equivalent. There is a clear sense that the algebraic category or theory in this case is more primitive or fundamental than the geometric counterpart, namely the sense that the description of the latter category is dependent on the former. As I have explained, nilpotent algebras are natural and well-behaved quotient algebras in the algebraic category. In contrast, their dual objects do not have any intrinsic descriptions (unless we are willing to abandon classical logic) but described externally by reversing the arrows pointing to and from nilpotent algebras. To reason with the geometric category, we have to appeal to the algebraic category, which is exactly what I have been doing so far (see Footnote \ref{derivative} for example). Thus, while it is true that every category has a dual category, it is often the case that one of these categories is more fundamental than the other if it has a (more) viable intrinsic description, and thereby more eligible for a realistic interpretation.

\subsection{Multi-fields algebras}

Nilpotent algebras or synthetic differential geometry, while having its own distinct advantages, does not aim to solve the interpretative problem of the scalar field that troubles the formalism of Einstein algebras. \emph{Multi-field algebras} proposed by Chen and Fritz ([2021]), which they simply called `field algebras', purport to address this problem.  In other words, the formalism of multi-field algebras aims at  (1) not privileging a particular type of physical fields that play the role of spacetime, but treating all physical fields on a par; (2) committing to only physical fields recognized by physics in its basic ontology. In this section, I will provide a relatively non-technical presentation of their approach with an emphasis on its philosophical aspects.

Since the aim is to not privilege a scalar field and to treat all physical fields on a par, we let the basic algebraic structure consist of all physical fields recognized by current physics such as fermion fields and a metric field, rather than (just) the scalar field. In this approach, a model for a (classical) field theory can be written down as $\{\Psi,\Phi,\Xi,...\}$, where $\Psi,\Phi,\Xi$ stand for different physical fields and each of them consists of all possible field configurations (together with dynamical information). Interpreting it realistically, the ontology consists of all field configurations of all fundamental physical fields, and the ideology consists of algebraic relations between field configurations of the same fields as well as those of different fields.\footnote{To address the concern of an extravagant ideology, Chen and Fritz tried to appeal to only algebraic relations that are useful for physics; see also Footnote \ref{Pl}.} One may object that this seems less parsimonious than an Einstein algebra containing just the scalar field. But to do physics, we still need to posit other fields in the Einstein-algebra framework, which is thereby no more parsimonious than a multi-field algebra. Since the formalism of multi-field algebras  deals away with a ghostly scalar field and the implausible dependence on it, it is more ontologically parsimonious and perspicuous.

There are two technical obstacles for develop multi-field algebras. First, as I have explained in Section 3.1, it is difficult to replace the scalar field by other fields. Secondly, different physical fields have diverse mathematical forms. For example, the electromagnetic field is a one-form, the fermion fields are spinor fields, and the metric field is a tensor field. So how can we put them all into one algebraic structure? What should be the algebraic operations that apply to different fields? 

These obstacles are overcome by the technique of natural operators in category theory, which is a convenient tool for constructing algebraic structures (Kolar et al. [1993]). This is roughly how it works. We first conceptualize a physical field as a functor $\mathcal{F}$ (equipped with a Lagrangian encoding its dynamical information). In pre-algebraic terms, a field functor is a functor from a customizable category of `spacetime' (in which the objects are certain representations of spacetime such as manifolds and the maps are certain transformations between them such as diffeomorphisms) to category of sets, assigning to every spacetime a set of field configurations. It follows from the definition of a functor that the field functor commutes with diffeomorphisms (or morphisms in general) between spacetimes, which is desirable because a physical field should be diffeomorphically invariant. (This is why fields should be conceptualized as functors in the first place.) For example, consider the category consisting of all coordinate representations of Minkowski spacetime as its objects and all Lorentz transformations as its maps. A real-valued scalar field, then, is a functor that specifies a smooth function for every coordinate representation of Minkowski spacetime that commutes with Lorentz transformation.\footnote{Note that in this approach I am still talking about scalar fields because the example of a scalar field is a simple heuristic example frequently used in physics. The problem of Einstein algebras is not involving scalar fields per se, but that the scalar field is indispensable to the formalism, i.e., it does not give a recipe of how to generalize the formalism  to different fields. } This way, Lorentz invariance is built into the description of the field, namely that the scalar field is not represented by a single smooth function but all of them related by Lorentz transformations.

Suppose the physical fields of interest include fields of different formal types like a spinor field and the electromagnetic field (a one-form)---the theory that studies them is called `spinor electrodynamics'. In order to include them in the same algebraic structure, we need to define multi-sorted algebraic operators that apply to both of them. It turns out that we can define such an algebraic structure using natural operators on the field functors. (Natural operators are just like natural transformations except that the latter are binary while the former can be $n$-ary.  A natural transformation is like a `higher-order' morphism between functors that preserves the functors' behaviour.) 
For example, we have a natural operation that maps a spinor field configuration $\Psi$ and an electromagnetic field configuration $A$ to a new spinor configuration $\sigma^\mu A_\mu \Psi$ (see Chen and Fritz [2021], p.197 (5.15)). 

If we list all the natural operators on those fields that are relevant to their physics, then we arrive at the desired multi-field algebra, according to which we can reconceptualize the field configurations as its simple elements rather than functors from a category of spacetime.\footnote{It is worth noting that what natural operators we find depends on the details of what functors are used to represent various fields. For any given physical field, there is a great deal of flexibility in customizing its functor in order to achieve the desired invariance of the field. For example, the electromagnetic field (represented by a 1-form tensor field) is invariant under gauge transformation, that is, adding the derivative of a scalar field to the electromagnetic field does not correspond to any physical change (for example, see Healey [2007] ). Accordingly, we can formulate functors in a way that identifies field configurations that are related by gauge transformations: instead of assigning all possible configurations of 1-form to a given manifold, we should assign only the quotient set of those configurations that does not distinguish between ones differing only by the derivative of a scalar field. This affects what natural operations we can define on the field.}


To take stock, unlike the formalism of Einstein algebras,  the elements in the field algebra are actual physical fields with energy and momentum and physical interactions.\footnote{Natural operations are abundant and some may not correspond to physical reality.  If we do not pare down natural operations to those that are required by physics, a homomorphism from one field algebra to another that preserves all physically-required operations but not all natural operations can generate distinct models that presumably does not correspond to genuine physical difference. 
	
	To give an example of a physically required operation: in order to formulate the Lagrangian of a scalar field $\phi$ according to a particular theory (such as $\phi^{4}$-theory), we need an operation on the scalar field that gives the `length' of the gradient of $\phi$, and this operation is one of the natural operations that characterize the field algebra, of which $\phi$ is an element. (According to the massless scalar $\phi^{4}$-theory in d dimensions---a commonly used toy example in physics---the Lagrangian density is given by $\mathcal{{L}}^{\mathrm{scalar}}(\phi)=\left(\frac{1}{2}g^{\mu\nu}\,(\partial_{\mu}\phi)\,(\partial_{\nu}\phi)-\frac{\lambda}{4!}\phi^{4}\right)\sqrt{|\det g|}$. Here, the commutative binary operation $\langle d-,d-\rangle$ on $\phi\times\phi$ is a natural operation that gives us the `length' of the gradient of $\phi$.)  
	
	But there are a vast number of natural operations on the field that are not required by physics. For example, every smooth operation is a natural operation on a scalar field (for every $n$-argument smooth function on $\mathbb{R}$, there is a corresponding smooth operation on a $C^\infty$-ring that takes its $n$ elements to another element). But  what we need for physics up to general relativity is arguably only \textit{commutative rings}, equipped with addition and multiplication (Geroch [1972]). } There is no tacit arena that physical fields live on, and the dynamical interactions of physical fields defined by natural operations are treated as primitive. Thus, it is a better implementation of relationalism than Einstein algebras.\footnote{It is worth noting that this approach is still very new and its foundational adequacy is yet to be explored. For one thing, for most field algebras, no complete list of natural operations required by physics is given (even the mathematical study of natural operations is relatively new). Moreover, Chen and Fritz ([2021]) noted that very little is known about how to do physics within this framework:
	
	\begin{quote}
		Our treatment of physics is limited to setting up the field equations, and we neither have a satisfactory theory of their solutions, nor do we have a field-algebraic treatment of Lagrangians and the principle of stationary action. Extending field algebra so as to incorporate these aspects of physics remains an open problem. (p.199) 
\end{quote}}

Unsurprisingly, the category of multi-field algebras is not dually equivalent to the category of manifolds, since the latter requires Rosenstock et al's constraints. It is also useful to note that field algebras are generally very different algebraic structures from all aforementioned algebraic structures (a multi-field algebra is generally equipped with more operators than $C^\infty(M)$). Note that even if in the hypothetical case of a multi-field algebra consisting of only scalar field configurations like Einstein algebras, they are still very different because the latter need to satisfy more constraints than multi-field algebras even in Geroch's original formalism (that is, besides those of Rosenstock et al).\footnote{An Einstein algebra has to satisfy certain constraints so that we can define other notions of differential geometry such as the metric tensor on it (to remind: the formalism of Einstein algebras allows us to do general relativity; see Geroch [1972], p.274). In contrast,  the `multi-field' algebra of smooth functions is subject to no such constraints because it is against the formalism of multi-field algebras to define the metric field or other fields through smooth functions. Instead, if the metric field is of fundamental interest, it should be directly included in the multi-field algebra.} So once again, a more interesting algebraic implementation of relationalism is not equivalent to manifoldism.

\subsection{Non-commutative algebras}

The final example of algebraicism I will mention involves non-commutative algebras (where multiplication is not commutative). This is related to an approach called `non-commutative geometry', which is an extension of the application of algebraic techniques in physics by lifting the constraint of commutativity standardly required of algebras of interest, and is of special interest for quantum physics (for example, see Connes and Marcolli [2008])\footnote{Non-commutative geometry is in fact a family of approaches to quantum geometry among the research programs for quantum gravity, but I will not delve into this topic.} This approach to algebraicism is worth mentioning because---to recall---one motivation for getting rid of manifolds mentioned by Geroch is to open new paths for quantum physics. Again I will stick to simplest examples and focus on their conceptual relevance.

I shall first note that commutativity is a default requirement on algebras in algebraicism if we implicitly think about physics in terms of manifolds or geometric spaces in general. 
This is because if we start with a manifold and a set of functions on the manifold and examine what algebraic relations are held among them, such functions are always commutative. 

But part of the spirit of algebraicism as a genuine alternative to manifoldism or substantivalism is precisely to examine what constraints inherited from manifoldism (or the geometric approach in general) are necessary and what are not. Thus it is natural to experiment with the absence of certain structures. It turns out that we can do a great deal of algebraic physics without the commutativity condition. In particular, we can still recover a part of differential geometry that matters for physics (in particular, general relativity and quantum field theory).  `Derivative'  is a particularly important notion in differential geometry (indeed it is the first notion to be re-defined based on Einstein algebras in Geroch [1972], p.272).\footnote{In (Geroch [1972]), a derivation $D$ is defined to be an automorphism on $C(M)$ that satisfies that for any ring elements $f,g$ (1)$D(f,g)=D(f)+D(g)$, (2) $D(fg)=D(f)g+fD(g)$, and (3) for any $x\in\mathbb{R}$, $D(x)=0$. These conditions parallel how `derivative' is defined in standard differential geometry.}  It turns out that this notion can survive the failure of commutativity because the defining feature, namely the Leibniz law, still makes sense without commutativity.\footnote{Let $\mathcal{A}$ be a non-commutative algebra and $\Omega$ be a bimodule over $\mathcal{A}$ (`(bi)module' is a generalization of `vector space'), which is to capture the notion of `one-form'. More technically, if $\mathcal{A}$ is an algebra, then $\Omega$ is a left module of it if there is a bilinear map $\mathcal{A}\times\Omega\to \Omega$ such that for any $a,b\in\mathcal{A}$ and $n\in\Omega$, $a(bn)=(ab)n$. A bimodule is both a left and a right module. Note that if $\mathcal{A}$ is commutative, modules are all bimodules. Then we can define a derivative to be a map $d$ from $\mathcal{A}$ to $\Omega$ such that $d(fg)=d(f)g+fd(g)$, which makes sense regardless of commutativity of $\mathcal{A}$.}

Why is non-commutativity related to quantum physics? We can understand the big-picture idea this way: a quantum field takes values not in definite quantities but operators, and operators do not necessarily commute. For example, the famous uncertainty principle in quantum mechanics says that the momentum operator and the position operators do not commute. Field values that are built out of such operators are thereby also not commutative. 

More detailedly, non-commutative geometry features a spectral triple $\langle \mathcal{A},\pi, D\rangle$, where $\mathcal{A}$ is a non-commutative algebra, $\pi$ is a representation function from $\mathcal{A}$ to a Hilbert space $\mathcal{H}$, and $D$ an operator on $\mathcal{H}$ known as a `Dirac operator' that roughly encodes the metric information (depending on what information to highlight, a spectral triple is also written as  $\langle \mathcal{A},\mathcal{H}, D\rangle$).\footnote{It might be worth noting that there is a close relationship between functions on a manifold and operators on a Hilbert space. For example, every smooth function $f$ defines an operator namely multiplying-by-$f$. All continuous functions on a compact Hausdorff space form a $C^*$-algebra just like all bounded operators on a Hilbert space (the latter is the canonical example of $C^*$-algebra). It might be of interest to readers that there are further connections between manifolds and Hilbert spaces made precise by topological quantum field theory (see Baez [2001], especially p.187, 192).}  

To illustrate, let's consider a simple example.  Let Hilbert space $\mathcal{H}$ consist of complex-valued and square-integrable functions on the real line (which can represent the wavefunction of a particle in a one-dimensional space), and Let $x, p$ respectively be the position and the momentum operator on $\mathcal{H}$, which obeys the canonical commutation relation between the two operators following the uncertainty principle, namely $[x,p]=i$ (setting the Planck constant $\hbar$ to 1).\footnote{For any $\phi\in \mathcal{H}$, $x:\mathcal{H} \to \mathcal{H}$ maps $\phi$ to $x\phi$, and $p:\mathcal{H}\to \mathcal{H}$ maps $\phi$ to $i\frac{d}{dx}\phi$.  A technical caveat: $x$ and $p$ are not bounded operators on Hilbert space and therefore do not satisfy the definition for spectral triples. To fix this, we can take the operators to be $e^{ip}$ and $e^{ix}$ instead. But we do not need to be concerned with this detail.
	
For any $\phi \in\mathcal{H}$, we have $(xp-px)\phi=xi\frac{d}{dx}\phi - (i\phi +xi\frac{d}{dx}\phi)=i\phi$
}  Let $\mathcal{A}$ be the algebra of polynomials in these operators over the field of complex numbers. It is clear that the this algebra is non-commutative given the commutator between $x$ and $p$. 

If the algebra in a spectral triple were commutative, the spectral triple would be equivalent to a spin manifold due to Connes' reconstruction theorem (Connes [2013]).\footnote{Here's a simple example of a commutative spectral triple: let the algebra in question be the algebra of all smooth functions on $\mathbb{R}$ which is commutative, let it be represented by the set of operators on $\mathcal{H}$ with each smooth function $f$ mapped to the operator multiplying-by-$f$.} But given that $\mathcal{A}$ is non-commutative, the spectral triple is not equivalent to any manifold-theoretic structure, and indeed any structure that is based on point sets, because we cannot reconstruct spacetime points from $\mathcal{A}$ (see also Huggett et al [2021], Bain [2006]). Here's why. As mentioned before, for the geometric dual object of an algebra to contain a point in the geometric category, there needs to be a homomorphism from the algebra to the field of complex numbers or real numbers, which is dual to a point. But it is clear that there is no such homomorphism from a $\mathcal{A}$ to $\mathbb{C}$ because the latter is commutative.\footnote{More detailedly, suppose for reductio there is such a homomorphism $f:\mathcal{A} \to\mathbb{C}$ that maps $x$ to $c$ and $p$ to $c'$. Then, since $[x,p]=i$, we have $f(xp)-f(px)=cc'-c'c=f(i)$. But this cannot be the case because $f(i)=i$ while $cc'=c'c$ ($\mathbb{C}$ is commutative).  Therefore there is no such homomorphism. } 

But perhaps unlike the commutative case, a point is no longer algebraically represented by a homomorphism from the algebra to $\mathbb{C}$ (or $\mathbb{R}$) in the non-commutative case. Perhaps it should rather be the maximal ideal of $\mathcal{A}$ (see Section 2), since this is another definition of `point' in the commutative case, and the notion of a maximal ideal is still well-defined for non-commutative algebras. However this does not work because the non-commutative algebra $\mathcal{A}$ (which is a Weyl algebra) does not have any non-zero maximal ideals (a zero ideal includes only zero---the quotient algebra it defines is just the whole algebra; algebras that have only zero ideals are called `simple') (see Coutinho [1997]). If we consider the zero ideal or equivalently, the entire algebra as corresponding to a point, not further divisible into subregions, the resulting geometrical picture is certainly very different from manifoldism. In the category of smooth manifolds (or other geometric structures), there is only one point object (zero-dimensional space) up to isomorphism. But the algebraic category in question contains many non-isomorphic non-commutative algebras, the dual objects of which cannot all be points.

There is a good amount of work in the literature exploring many different applications of non-commutative geometry to physics (for instance, see Connes and Marcolli [2008])---for example, one claim is that we are able to do quantum field theory with simpler reasoning with Lagrangians and actions. But we can appreciate the conceptual value of this approach at the general level without looking into these details.  Since we cannot reconstruct manifolds from non-commutative algebras, this approach allows us to generalize beyond the manifold-theoretic approach and broadens the landscape of possibilities while still being able to do physics. In other words, non-commutative geometry shows that positing a common arena for physical fields such as a manifold amounts to an unnecessary theoretical constraint in addition to its ontological weight: that the field configurations have to commute. Such a generalization is much needed because of the need for new physics and to solve a multitude of difficulties of quantum gravity arising in the manifold-theoretic framework. Many have speculated that we do not have a physically well-defined notion of regions smaller than the Planck scale. Non-commutative geometry provides a possible way of implementing this idea other than considering spacetime as a discrete lattice.

Let me rehearse a point made in the beginning of Section 3, since at this point readers might again feel that we haven't really dispensed with ordinary geometry or manifolds given that we still refer to them while constructing the spectral triple ($\mathcal{A}$ and $\mathcal{H}$ in particular). It is worth emphasizing that the algebra in question can be described completely abstractly without referring to particular operators on the Hilbert space (namely $C^\star$-algebra), and the Hilbert space can also be defined abstractly up to isomorphism. We have not smuggled manifolds in.  

\subsection{Problem of pre-geometry and the constructive program}

Finally, I would like to briefly comment on two standard challenges for relationalism: (1) the problem of pre-geometry: how is it even possible to dispense with a common arena for physical ongoings (see Norton [2008])?\footnote{The term `problem of pre-geometry' is coined by Linnemann and Salimkhani [2021] which I find helpful in describing Norton's [2008] concern.} (2) The constructive program: if spacetime does not exist at the fundamental level, how can we construct it in order to account for the manifestly chrono-geometric world? 

\textit{Ad} (1), the fact that we can do physics (to some extent) in algebraicism as a genuine relationalist view implies that we can in fact dispense (to the same extent) with spacetime as a background arena (see also Menon [2019]). For example, Norton asked how we can account for spacetime coincidence without a spacetime. Take multi-field algebraicism as an example.  Consider two configurations $f,g$ of any physical field, which can be respectively realized as a function with value $n$ at point $p$ (and negligible elsewhere) and a function with value $m$ at $p$ (and negligible elsewhere).  let $h_1$ be the element that can be realized as having value $m+n$ at $p$ (and negligible elsewhere), and $h_2$ be the one realizable as having $n$ at $p$ and $m$ at $q$ (and negligible elsewhere). The spacetime coincidence of the peaks of $f, g$ is encoded in the fact that the addition of $f$ and $g$ is $h_1$ rather than $h_2$. Indeed, such a response to Norton is not new, and in fact belongs to a standard line of response from relationalists (see Pooley [2013]).\footnote{As Pooley  ([2013], p.29) pointed out, `the most natural	is to take spatiotemporal coincidence as primitive (as many relationalists have
	done;	e.g.,	Rovelli	(1997,	194))'. The algebraic response I outline is one concrete way of implementing this general strategy.}

\textit{Ad} (2), in algebraicism, in general, geometric notions are `recovered' from the algebraic objects simply via categorical duality. For example, the spatial parthood relations are interpreted as surjective homomorphisms between algebras. In this way some geometric claims can be made true by the corresponding algebraic relations between fields. It is worth emphasizing that there is a clear sense that geometry is derived from algebras in algebraicism which is that the geometric category dual to the algebraic category of interest often lacks an intrinsic description (see Section 4.1). Note that categorical duality does not directly address the problem of how our ordinary claims about space and time are made approximately true by the underlying category, which may be very different from our phenomological spacetime. This, however, is a separate substantial question in physics regarding how ordinary spacetime emerges from a more fundamental physical theory, which is out of the scope here.\footnote{As an example among many, there is ongoing research on how ordinary continuous spacetime can emerge from the `causal set' that consists of partially ordered discrete events.} 
		
		
\section{Conclusion}
In this paper, I have defended algebraicism---an algebraic implementation of relationalism according to which physical fields exist fundamentally without spacetime and can be understood in terms of their structural relations---as a genuine relationalist view alternative to manifoldism or substantivalism according to which spacetime exists fundamentally. I have argued that the standard version of algebraicism that is considered equivalent to manifoldism is not a particularly interesting or attractive version of algebraicism. Instead, I have presented three examples of algebraicism that are demonstratively not equivalent to manifoldism or substantivalism and have their own conceptual advantages. In particular, nilpotent algebras can account for vectorial quantities. Multi-field algebras deals away with the dependence on a ghostly scalar field and are thereby more ontologically parsimonious and perspicuous. Non-commutative algebras are amenable to quantum physics.  

Algebraicism is a foundational promise to generalize manifoldism for the need of physics and other theoretical considerations, rather than being merely a representational alternative  that  neither matters for empirical consequences nor for our fundamental ontology. I thereby extends an invitation to readers to go beyond the equivalence claim and explore the vast potential of this approach.

\end{document}